# The Comparative Trap: How Social Comparison Orientation Drives Problematic Generative AI (GenAI) Use

Xuchao Zhang , Jihye Lee

School of Design, Pusan National University, Busan, South Korea

ABSTRACT: Although Generative AI (GenAI) improves task efficiency in the short term, it creates competitive pressures that perpetuate individuals' fear of being eliminated, thereby increasing the risk of problematic use. Existing research has focused on the perspective of individual psychological vulnerability, but has neglected the social comparison context caused by GenAI. This study examines the direct effects of social comparison orientation on problematic GenAI use and explores their indirect effects via emotional and cognitive mechanisms, grounded in the Person-Affect-Cognition-Execution (I-PACE) model. The research analyzed data from 396 Chinese GenAI users using SEM and bootstrap methods. Findings show that social comparison orientation has a significant direct impact on problematic GenAI use and can additionally influence AI flow and perceived irreplaceability through fear of missing out (FoMO), finally leading to problematic GenAI use.

## 1. Introduction

Generative AI (GenAI) can significantly improve individuals' task efficiency in a short period of time (Desdevises, 2025), but long-term use may pose risks of loss of control. Studies show that frequent use of AI tools reduces deep thinking and may even impair memory (Georgiou, 2025). In addition, although GenAI, as a productivity-enhancing technology, significantly accelerates task completion and improves output quality (Noy & Zhang, 2023), the social competitive pressure and anxiety about falling behind it are also intensifying (Broadfoot & Rockey, 2025; Dinh, 2025). In this context, GenAI is not only a tool for improving efficiency but also plays a role in expanding performance gaps and social comparison. Therefore, users also increase their collaboration with GenAI to maintain their own competitiveness. However, long-term reliance on a specific technology may gradually lead users to develop dependence (Zhang et al., 2024) and may even result in addiction risks (Kooli et al., 2025) or problematic use behaviors (Chen, 2025).

Existing research suggests that individuals with higher levels of anxiety and depression are more likely to develop psychological dependence on AI (Chen, 2025; Heng & Zhang, 2025; Huang et al., 2024). Low self-esteem (Yao et al., 2025), attachment (Huang & Huang, 2025), cognitive inertia (Ye et al., 2025), and self-efficacy (Zhang et al., 2024) have also been identified as essential antecedents of problematic GenAI use. Current studies mainly explain the formation of problematic GenAI use from the perspective of individual psychological vulnerability, while overlooking the social comparison attributes of GenAI (Broadfoot & Rockey, 2025; Dinh, 2025). Meanwhile, research on social comparison orientation has been mainly limited to social media contexts (Feinstein et al., 2013; Feng et al., 2025; Haferkamp & Krämer, 2011; Hou et al., 2023; Kalpidou et al., 2011; Kim et al., 2021; Kross et al., 2013; Lee, 2014; Reer et al., 2019; Servidio et al., 2021; Vogel et al., 2014). However, unlike the comparisons commonly observed in social media research, which focus on lifestyles, social status, or appearance, in GenAI contexts, comparisons are mainly centered on productivity and efficiency, and AI platforms have quietly become new sites for social comparison. The mechanisms underlying problematic use in the AI era may differ fundamentally from those observed in traditional social networking environments. Therefore, as a personal trait that activates social comparison, the impact of social comparison orientation on problematic GenAI use has not yet been adequately explored.

By addressing these gaps, this study is built on the Person-Affect-Cognition-Execution (I-PACE) model and tests a multiple-mediation structural equation model. Specifically, social comparison orientation, a vulnerability trait at the person level, may directly promote problematic GenAI use or operate through fear of missing out (FoMO) at the affect level, which is then transmitted to AI flow and perceived irreplaceability at the cognition level, ultimately driving problematic GenAI use at the execution level.

## 2. Literature Review

### *2.1. Person-Affect-Cognition-Execution (I-PACE) model*

The I-PACE model is a theoretical framework that explains problematic technology use behaviours (Brand et al., 2016, 2019) by emphasising interactions among variables at the person, affect, cognition, and execution levels (Zhang et al., 2024). Specifically, the person level includes susceptibility factors such as biophysical makeup, personality traits, psychopathology, social cognition, and motivations for use; the affect level refers to an individual's emotional response to environmental stimuli; the cognition level relates to cognitive biases in the technology use process; and the execution level encompasses the outcomes of problematic behaviours, such as an increased frequency of use and decreased control.

The I-PACE model has been used in existing studies to explain online gaming addiction (Jhone et al., 2021), smartphone addiction (Liu et al., 2025; Wanjiru et al., 2025; Mehmood et al., 2021), and problematic social media use (Kavakli, 2025) in digital contexts. Recently, studies have extended the I-PACE model to GenAI scenarios to reveal the psychological mechanisms underlying dependence and addictive tendencies in AI use, and have preliminarily verified the applicability of this model in explaining problematic use of emerging technologies (Chen, 2025; Du et al., 2026; Revesai, 2025; Ye et al., 2025). Compared with digital media, the instant generation and feedback of GenAI significantly shorten the distance between cognition and execution, allowing individuals' cognitive biases to be more rapidly transformed into problematic behaviors. Therefore, the I-PACE model is a theoretical framework well-suited to revealing the formation process of individuals' problematic behaviors in AI use contexts.

In addition to examining the direct impact of social comparison orientation on problematic GenAI use, this study uses the I-PACE model to shed light on the psychological mechanisms that underlie the development of problematic GenAI use. Specifically, social comparison orientation, as an individual trait at the person level, influences the affect level through FoMO and, through two pathways, affects the cognition level (AI flow and perceived irreplaceability), ultimately leading to problematic GenAI use at the execution level.

## 2.2. Social Comparison Orientation

Social comparison orientation is the tendency to habitually compare one's abilities and performance with those of others (Festinger, 1954; Gibbons & Buunk, 1999). According to social comparison theory, when clear evaluation criteria are lacking, individuals tend to obtain information about their own abilities and values by comparing themselves with others (Festinger, 1954). Therefore, social comparison orientation reflects the degree to which individuals rely on external social references during self-evaluation processes (Suls & Wheeler, 2013). Individuals with high social comparison orientation show sustained sensitivity to others' performance, and their self-concept is more likely to exhibit instability and uncertainty (Gibbons & Buunk, 1999; Guimond, 2006), and thus it is regarded as a relatively stable individual difference trait. Research in social media contexts shows that social comparison orientation directly predicts problematic smartphone use (Servidio et al., 2021) and social network addiction (Kim et al., 2021; Hou et al., 2023; Feng et al., 2025). Meanwhile, social comparison orientation can also indirectly influence problematic use outcomes through emotional and cognitive mechanisms such as FoMO (Reer et al., 2019), depression (Feinstein et al., 2013), self-evaluations (Haferkamp & Krämer, 2011), self-esteem (Kalpidou et al., 2011; Lee, 2014; Vogel et al., 2014), and well-being (Kross et al., 2013). Taken together, social comparison orientation plays a vital role in problematic technology use, which can be amplified by both direct effects and affective and cognitive processes.

However, existing research has not yet systematically investigated how social comparison orientation influences GenAI use. Unlike social platforms centered on information display and interaction (McComb et al., 2023; Ozimek et al., 2023), in the AI era, users commonly collaborate with GenAI to improve task efficiency and output quality, thereby changing the basis on which social comparison occurs (Noy & Zhang, 2023). That is, individuals tend to regard their mastery of AI and its output capabilities as part of their own abilities, forming an extended self (Litvinova et al., 2024), and shift toward comparisons of human–AI collaborative performance (Jansen et al., 2025).

Therefore, this study defines social comparison orientation as a person-level vulnerability factor and proposes that it may exert a direct effect on problematic GenAI use.

H1: Social comparison orientation positively influences problematic GenAI use.

### *2.3. Fear of Missing Out (FoMO)*

FoMO is a prevalent form of anxiety characterized by the concern that others are engaging in valuable or meaningful experiences in one's absence. This perception often leads to a strong compulsion to remain connected to others or stay online (Przybylski et al., 2013). Existing studies generally regard FoMO as an emotional response triggered by uncertainty, perceived potential loss, and social comparison pressure. A large body of empirical research has confirmed that FoMO is a key emotional predictor of problematic technology use, including problematic use of social media (Casale et al., 2018; Rozgonjuk et al., 2020; Fioravanti et al., 2021), smartphone (Elhai et al., 2018, 2020), and the internet (Stead & Bibby, 2017). Therefore, FoMO has been widely considered a mature and key emotional risk factor in research on problematic media use. In social media contexts, social comparison orientation is an essential antecedent of FoMO. Related research shows that individuals with higher social comparison orientation are more likely to be exposed to others' positive experiences, thereby intensifying concerns about missing out on important experiences (Reer et al., 2019). Further research has revealed that upward social comparison can indirectly influence a series of maladaptive behaviors through FoMO (Wang et al., 2023). In addition, FoMO is more appropriately conceptualized as an emotional outcome following social comparison rather than as a preceding variable (Burnell et al., 2019). Together, these findings indicate that FoMO is not an emotional state independent of the social comparison process. Still, rather a key emotional response activated under social comparison pressure, mediating between social comparison and problematic use behaviors.

Compared with social media, FoMO in GenAI use contexts shows more pronounced technology- and competition-oriented characteristics. In the GenAI era, mastering and efficiently using AI tools has gradually been regarded as an essential source of productivity and competitive advantage (Noy & Zhang, 2023). For individuals with a high social comparison orientation, when they perceive that their peers have significantly improved their work efficiency or output quality through human–AI collaboration, while they themselves may be at risk of technological obsolescence, such social comparison is more likely to transform into an AI-contextualised FoMO, centred on concerns about marginalisation or elimination (Méndez-Suárez et al., 2026). In other words, it is the fear of being marginalised or eliminated due to a failure to master AI tools. In GenAI use contexts, social comparison orientation serves as an essential psychological channel that transforms external technological competition pressure into internal emotional anxiety.

This study defines FoMO as an emotional variable at the affect level and considers it to play a key mediating role between the person and cognition levels (Brand et al., 2019). Therefore, in GenAI use contexts, individuals with higher social comparison orientation are more likely to experience higher levels of FoMO when facing ability gaps amplified by AI.

H2: Social comparison orientation positively influences FoMO.

### 2.4. AI Flow

Flow refers to a highly immersive cognitive state experienced by individuals when engaging in an activity (Csikszentmihalyi, 1990). When in a flow state, individuals focus intently on the current task and perceive time passing more quickly. Although flow is generally considered to be a positive experience, studies in the gaming (Li et al., 2021), offline activities (Tan et al., 2023), and social media (Brailovskaia & Margraf, 2024) have shown that FoMO can trigger problematic use behaviors by strengthening flow experiences. In other words, when individuals experience FoMO, flow is often used as a short-term psychological compensation to temporarily alleviate uncertainty and pressure.

In GenAI contexts, GenAI systems usually provide instant feedback and highly personalized interactions, which make users more likely to enter sustained interactions. In today's highly competitive environment enabled by GenAI, those with higher levels of FoMO are keen to interact with AI continuously. This is both to enhance their competitiveness and to reduce the risk of being eliminated. In this process, FoMO not only increases use motivation but also strengthens individuals' functional dependence on immersive experiences, making them more likely to enter AI flow during interactions with GenAI. Previous research has confirmed a significant association between flow and problematic use behaviours in contexts such as smartphone addiction (Chen et al., 2025; Jo & Baek, 2023), internet addiction (Khang et al., 2013), gaming addiction (Chou & Ting, 2003; Footitt et al., 2024; Putra et al., 2024), and problematic social media use (Brailovskaia et al., 2018; Brailovskaia & Margraf, 2024; Miranda et al., 2023).

This study's AI flow is defined as a variable at the cognition level and is considered an essential mediating process that transforms affect-level states into problematic use at the execution level. When AI flow is motivated by FoMO, the resulting experience is more likely to weaken individuals' self-control, increasing the risk of problematic GenAI use.

H3: FoMO positively influences AI flow.

H4: AI flow positively influences problematic GenAI use.

### 2.5. Perceived Irreplaceability

Perceived irreplaceability is when people think a product or service is unique and can't be replaced (Grau, 2004; Liu & Wang, 2023). Individuals who regard it as the only means to satisfy their needs are more likely to develop dependent usage patterns, and FoMO is a key emotional factor that drives such patterns (Beyens et al., 2016; Jiang, 2016; Wong et al., 2015). Further research has confirmed that FoMO strengthens individuals' perceived irreplaceability of platforms, thereby leading to social media addiction behaviors (Cao, Yu, et al., 2025). Furthermore, studies have demonstrated a strong link between perceived irreplaceability and problematic usage behaviours in various contexts, such as online health platforms (Pedeliento et al., 2016), smart medical wearables (Chau et al., 2019), brand relationships (Sreejesh, 2015), and social media use (Cao, Gao, et al., 2025).

In highly competitive social environments, individuals with strong FoMO tend to view AI as a means of maintaining a competitive advantage, fearing being left behind. It has been found that high FoMO individuals still show a clear preference for AI, despite the fact that AI output may not be accurate enough (Kreps et al., 2023), This not only reflects their fear of being rendered obsolete by the technology but also suggests that their reliance on AI goes beyond the functional level and is accompanied by a deep psychological identity (Li, 2024). As a result, individuals gradually come to see GenAI as an indispensable means of survival.

This study defines perceived irreplaceability as a cognition level variable and posits that it mediates between the affect and execution levels. In GenAI contexts, FoMO is not only a psychological state but also significantly increases the risk of problematic GenAI use by strengthening individuals' perceived irreplaceability of the technology. The following hypotheses and model are therefore proposed (Figure 1).
H5: FoMO positively influences perceived irreplaceability.
H6: Perceived irreplaceability positively influences problematic GenAI use.

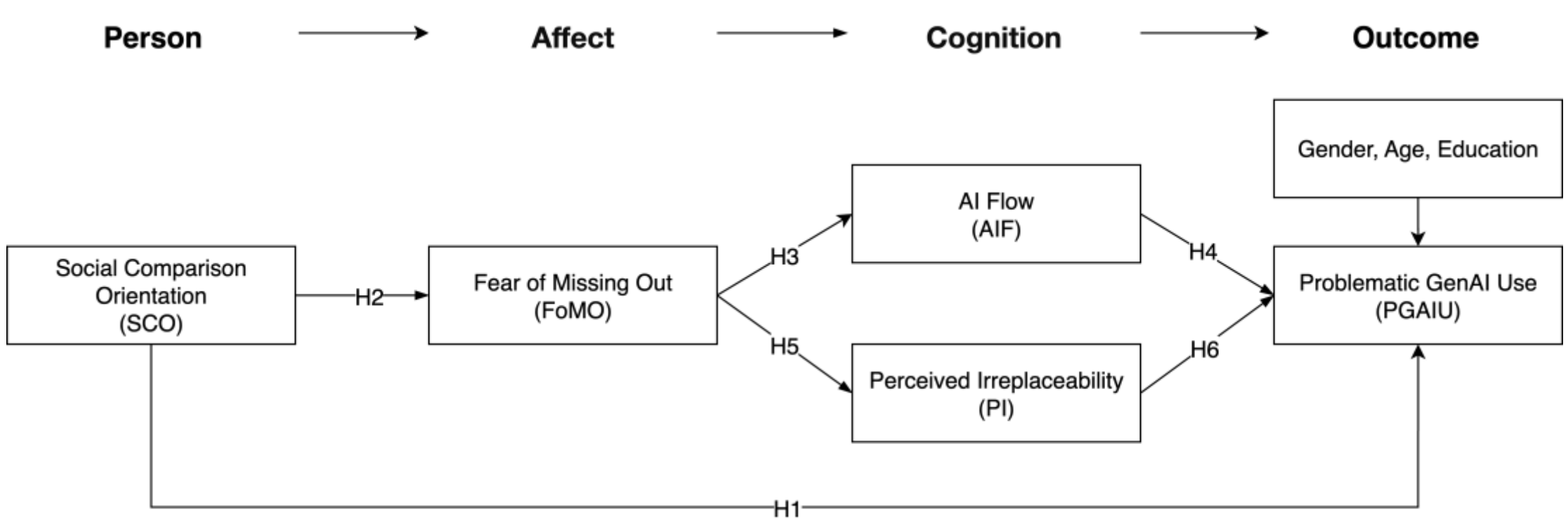

**Figure 1.** The research model.

## 3. Method

### *3.1 Participants*

The survey was anonymously distributed and collected 526 questionnaires from Chinese GenAI users. To encourage questionnaire completion, respondents received a 3 RMB reward. To ensure data quality, screening items and attention check questions were included. After removing 130 responses with no GenAI use experience and low-quality responses, 396 valid samples were retained (Table 1). Of the respondents, 48.5% male and 51.5% female. 97% aged 19-35. 86.1% held a bachelor's degree or above. Consistent with this distribution, core GenAI users are mainly young individuals with higher levels of education (Lane, 2024). Therefore, the sample of this study is highly representative and explanatory.

**Table 1**. Demographic information of the participants.

| | Category | Frequency | Percentage （%） |
|---|---|---|---|
| Gender | Male | 192 | 48.5 |
| | Female | 204 | 51.5 |
| Age | 19–23 | 168 | 42.4 |
| | 24–27 | 136 | 34.4 |
| | 28–31 | 63 | 15.9 |
| | 32–35 | 17 | 4.3 |
| | ⩾36 | 12 | 3.0 |
| Education | Associate degree or below | 55 | 13.9 |
| | Undergraduate | 253 | 63.9 |
| | Master | 72 | 18.2 |
| | PhD | 16 | 4.0 |
| Total | | 396 | 100 |

### 3.2. Instruments Development

The questionnaire's first part introduced the research background, research objectives, and informed consent information to the respondents. The second part collected participants' demographic information. The third part included five core constructs measured by 21 items. To ensure content validity, all measurement items were adopted from established scales used in previous studies and adapted for use in the context of GenAI. All items were measured using a seven-point Likert scale. Social comparison orientation was adapted from Gibbons and Buunk (1999); FoMO was adapted from Przybylski et al. (2013) and Chen (2025); AI flow was adapted from Hsu and Lu (2004) and Yao et al. (2025); Perceived irreplaceability was adapted from Subramani (2004) and Cao, Yu, et al. (2025); and problematic GenAI use was adapted from Hu et al. (2023). Before formal data collection, a pilot test was run with 50 users with GenAI experience. The results indicated that respondents consistently understood the item wording and that it was suitable for subsequent large-scale data collection.

## 4. Results

### 4.1 The Measurement Model

The measurement model was assessed before estimating the structural model. According to Schumacker and Lomax (2008) criteria, all items in this study had factor loadings greater than 0.500. Cronbach's alpha coefficients were all above 0.800. Convergent validity was assessed using CR and AVE. Prior research indicates that AVE values above 0.50 indicate sufficient indicator convergence (Fornell & Larcker, 1981). This criterion is met in the measurement model (Table 2). According to Chin (1998), discriminant validity is present when the square root of each AVE is greater than the related correlation coefficient. The measurement model met this standard (Table 3). We checked for multivariate normality. Mardia's multivariate kurtosis coefficients for all models exceeded the recommended threshold of 5, indicating significant departures from multivariate normality (Byrne, 2010). When data are not normal, Maximum Likelihood estimation can be biased, leading to a higher chi-square statistic and inaccurate parameter significance (Curran et al., 1996; Gao et al., 2008). To address this, we employed the Bollen–Stine bootstrap procedure to get more reliable model-fit estimates (Bollen & Stine, 1992; Enders, 2005). The results conformed to the thresholds proposed by Hu and Bentler (1999): $\chi^2/df = 1.27$, AGFI = 0.94, TLI = 0.99, CFI = 0.99, and RMSEA = 0.03 (Table 4).

**Table 2**. Results of construct validity and reliability analysis.

| Variable | Item | Mean | Std. Dev | Loading | α | CR | AVE |
|---|---|---|---|---|---|---|---|
| SCO | SC1 | 4.15 | 1.46 | .749 | .875 | .876 | .639 |
| | SC2 | | | .790 | | | |
| | SC3 | | | .802 | | | |
| | SC4 | | | .852 | | | |
| FoMO | FoMO1 | 3.87 | 1.44 | .669 | .878 | .851 | .535 |
| | FoMO2 | | | .713 | | | |
| | FoMO3 | | | .635 | | | |
| | FoMO4 | | | .900 | | | |
| | FoMO5 | | | .896 | | | |
| AIF | AIF1 | 4.94 | 1.21 | .737 | .836 | .841 | .570 |
| | AIF2 | | | .799 | | | |
| | AIF3 | | | .747 | | | |
| | AIF4 | | | .736 | | | |
| PI | PI1 | 4.41 | 1.54 | .846 | .899 | .901 | .752 |
| | PI2 | | | .911 | | | |
| | PI3 | | | .842 | | | |
| PGAIU | PGAIU1 | 4.25 | 1.44 | .825 | .883 | .885 | .606 |
| | PGAIU2 | | | .836 | | | |
| | PGAIU3 | | | .752 | | | |
| | PGAIU4 | | | .704 | | | |
| | PGAIU5 | | | .768 | | | |

*SCO* Social Comparison Orientation, *FoMO* Fear of Missing Out, *AIF* AI Flow, *PI* Perceived Irreplaceability, *PGAIU* Problematic GenAI Use

**Table 3.** Discriminant validity of the measurement model.

| Constructs | SCO | FoMO | AIF | PI | PGAIU |
|---|---|---|---|---|---|
| SCO | **.799** | | | | |
| FoMO | 0.498 | **.731** | | | |
| AIF | 0.202 | 0.346 | **.755** | | |
| PI | 0.410 | 0.515 | 0.431 | **.867** | |
| PGAIU | 0.457 | 0.517 | 0.593 | 0.591 | **0.778** |

### *4.2. The Structural Model*

SEM metrics showed good model fit: $\chi^2/df = 1.21$, AGFI = 0.93, TLI = 0.99, CFI = 0.99, RMSEA = 0.02 (Table 4). The test results supported all six hypotheses (Table 5). Social comparison orientation is positively associated with problematic GenAI use ($\beta = 0.241$, $p < .001$). It is also positively associated with FoMO ($\beta = 0.545$, $p < .001$). These results support hypotheses H1 and H2. FoMO is positively associated with AI flow ($\beta = 0.423$, $p < .001$), and AI flow is positively associated with problematic GenAI use ($\beta = 0.486$, $p < .001$), supporting hypotheses H3 and H4. FoMO is positively associated with perceived irreplaceability ($\beta = 0.591$, $p < .001$), and perceived irreplaceability is positively associated with problematic GenAI use ($\beta = 0.367$, $p < .001$), supporting hypotheses H5 and H6 (Figure 2).

**Table 4.** Bollen–Stine–corrected model fit indices.

| Model | Bollen-Stine $\chi^2$ | $\chi^2$/DF | AGFI | TLI | CFI | RMSEA |
|---|---|---|---|---|---|---|
| Measurement model | 226.64 | 1.27 | .94 | .99 | .99 | .03 |
| Research model | 291.15 | 1.21 | .93 | .99 | .99 | .02 |
| Recommended criteria | p > .05 | < 5.0 | > .90 | > .90 | > .90 | < .08 |

**Table 5.** The results of the hypothesis test.

| Hypotheses | Hypothesized path | B | β | S. E | t | Result |
|---|---|---|---|---|---|---|
| H1 | SCO → PGAIU | .244 | .241 | .046 | 5.307*** | Supported |
| H2 | SCO → FoMO | .511 | .545 | .059 | 8.669*** | Supported |
| H3 | FoMO → AIF | .375 | .423 | .054 | 6.884*** | Supported |
| H4 | AIF → PGAIU | .593 | .486 | .063 | 9.365*** | Supported |
| H5 | FoMO → PI | .682 | .591 | .069 | 9.837*** | Supported |
| H6 | PI → PGAIU | .344 | .367 | .043 | 7.976*** | Supported |

*SCO* Social Comparison Orientation, *FoMO* Fear of Missing Out, *AIF* AI Flow, *PI* Perceived Irreplaceability, *PGAIU* Problematic GenAI Use

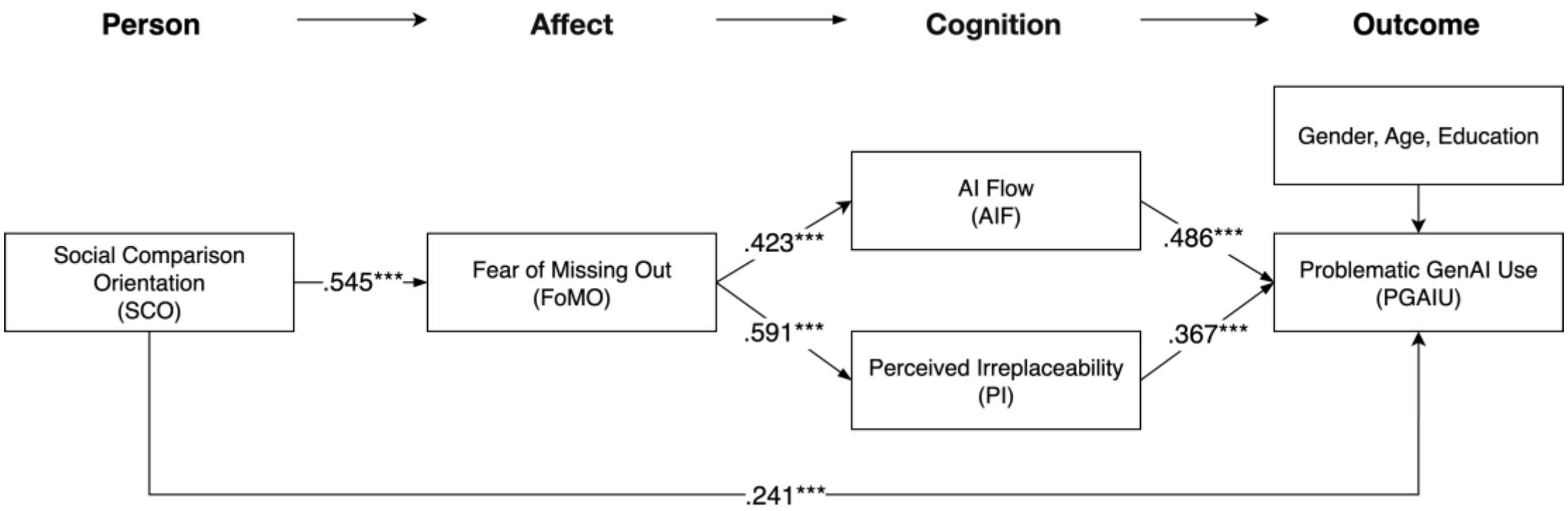

**Figure 2.** The research model with its standardised coefficients.

### 4.3. Direct, Indirect, and Total Effects

Mediation effects were assessed using a bootstrap procedure. The findings suggest that a social comparison orientation has a notable indirect effect on problematic use of GenAI. On the one hand, social comparison orientation influences AI flow through FoMO, which, in turn, leads to problematic GenAI use (β = 0.113, 95% CI [0.073, 0.170]). On the other hand, social comparison orientation influences perceived irreplaceability through FoMO, which, in turn, leads to problematic GenAI use (β = 0.120, 95% CI [0.075, 0.189]). The difference between the two indirect paths via AI flow and perceived irreplaceability is not significant (β = −0.006, 95% CI [−0.083, 0.066]).

Overall, the direct effect of social comparison orientation on problematic GenAI use is also significant (β = 0.244, 95% CI [0.136, 0.360]), and the total indirect effect is significant as well (β = 0.233, 95% CI [0.170, 0.316]). Therefore, the effect of social comparison orientation on problematic GenAI use in this model follows a partial mediation pattern. Finally, the total effect of social comparison orientation on problematic GenAI use is significant (β = 0.478, 95% CI [0.361, 0.594]) (Table 6).

**Table 6.** Bias-corrected bootstrap estimates of direct, indirect, and total effects.

| Effect | β | SE | z | Bias-corrected 95% CI |
|---|---|---|---|---|
| Specific indirect effects | | | | |
| SCO → FoMO → AIF → PGAIU | 0.113 | 0.024 | 4.708 | [0.073, 0.170] |
| SCO → FoMO → PI → PGAIU | 0.120 | 0.028 | 4.286 | [0.075, 0.189] |
| Difference (Flow - PI) | -0.006 | 0.037 | -0.162 | [-0.083, 0.066] |
| Overall effects | | | | |
| Direct Effect | 0.244 | 0.057 | 4.281 | [0.136, 0.360] |
| Total Indirect Effect | 0.233 | 0.037 | 6.297 | [0.170, 0.316] |
| Total Effect | 0.478 | 0.059 | 8.102 | [0.361, 0.594] |

*SCO* Social Comparison Orientation, *FoMO* Fear of Missing Out, *AIF* AI Flow, *PI* Perceived Irreplaceability, *PGAIU* Problematic GenAI Use, *Bootstrap samples = 5000, Bias-corrected 95% CI*

## 5. Discussion

### 5.1. Result discussion

Based on the I-PACE model, this study examined how social comparison orientation drives problematic GenAI use through affective and cognitive processes. The results show that social comparison orientation not only indirectly triggers problematic GenAI use through affect and cognition pathways, but also directly affects it.

First, social comparison orientation positively predicts problematic GenAI use (H1). This finding is consistent with prior studies on the role of social comparison orientation in problematic social media use and smartphone addiction (Kim et al., 2021; Servidio et al., 2021), and extends these findings to the GenAI context. Notably, this study finds that social comparison orientation remains directly significant after introducing FoMO and cognition variables. In GenAI environments, individual ability is no longer determined solely by knowledge and skills, but is closely related to the level of human–AI collaboration. For individuals with higher social comparison orientation, such human–AI performance comparisons are more likely to be incorporated into self-evaluation processes, keeping individuals continuously focused on their own and others' AI mastery and increasing the risk of problematic GenAI use. Overall, as a key vulnerability trait at the person level of the I-PACE model, social comparison orientation demonstrates independent and stable explanatory power in GenAI contexts.

Second, social comparison orientation positively predicts FoMO (H2). This result aligns with previous social media studies indicating that social comparison orientation induces FoMO (Reer et al., 2019), and with the theoretical expectation that FoMO is an affect-level emotional variable in the I-PACE model (Brand et al., 2019). However, the FoMO identified in this study differs from FoMO in social interaction or life experiences, and instead reflects anxiety about missing out on productivity, efficiency, and technological advantages. When individuals perceive that others achieve significant performance gains through human–AI collaboration, those with higher social comparison orientation are more likely to internalize such gaps as concerns about marginalization or elimination, thereby continuously transforming social comparison pressure into emotional tension.

Third, FoMO influences problematic GenAI use through two complementary cognitive pathways. On the one hand, FoMO positively affects AI flow (H3), and AI flow, in turn, positively predicts problematic GenAI use (H4). This result is consistent with prior findings showing that FoMO promotes problematic media use by strengthening flow experiences (Brailovskaia et al., 2018; Jo & Baek, 2023), and indicates that in GenAI contexts, immersive experiences may serve as a short-term psychological strategy to cope with competitive anxiety while weakening self-regulation in the long term. On the other hand, FoMO significantly increases perceived irreplaceability (H5), which, in turn, positively predicts problematic GenAI use (H6). This is consistent with prior research on the role of perceived irreplaceability in social media dependence and continued use of online platforms (Cao, Yu, et al., 2025), and indicates that when AI is perceived as the only tool for completing learning or work tasks, its functional value is constructed as indispensable, leading individuals to abandon independent thinking during decision-making and develop highly AI-dependent work patterns. At the same time, the indirect effects of the two paths via AI flow and perceived irreplaceability do not differ significantly in magnitude. That is, these two paths reflect cognitive mechanisms centered on experiential reinforcement and functional dependence that jointly drive the formation of problematic GenAI use. Unlike prior studies that emphasize the dominant role of immersion in problematic media use, in GenAI contexts, dependence on AI for task completion plays an equally important role in promoting problematic use.

## 5.2. Theoretical implications

This study provides three critical theoretical implications. First, based on the I-PACE model, this study systematically reveals the psychological mechanisms through which social comparison orientation drives problematic GenAI use, thereby extending the theoretical perspective of existing research on problematic GenAI use. Prior studies have mainly explained the formation of problematic GenAI use from the standpoint of individual psychological vulnerability, such as anxiety and depression (Chen, 2025; Heng & Zhang, 2025; Huang et al., 2024), low self-esteem (Yao et al., 2025), and attachment (Huang & Huang, 2025). This study further extends the explanation to the social competition intensified by GenAI. It indicates that problematic GenAI use is not only derived from psychopathological states but also from highly comparable technology use environments. Under efficiency-centered social evaluation standards, social comparison orientation, as a vulnerability trait at the person level of the I-PACE model, not only exerts indirect effects through affect and cognition, but also has an independent direct effect on problematic GenAI use, thereby extending the application boundaries of the I-PACE model in GenAI research (Brand et al., 2016, 2019).

Second, this study clarifies that FoMO is a key emotional mediator through which a social comparison orientation is transformed into problematic GenAI use. Previous studies in social media contexts have verified the critical role of FoMO in problematic use (Elhai et al., 2018; Reer et al., 2019; Rozgonjuk et al., 2020). However, these studies have conceptualized FoMO mainly as a form of social anxiety. This study conceptualizes FoMO as anxiety about missing productivity, efficiency, and technological advantages in GenAI use contexts. This finding not only confirms FoMO as a core emotional variable at the affect level within the I-PACE model, but also extends the concept of FoMO from social interaction contexts to GenAI use contexts.

Third, this study reveals two cognitive pathways underlying problematic GenAI use. Prior studies explaining GenAI dependence have often focused on a single-path mechanism (Huang & Huang, 2025; H. Li, 2024; Zhang et al., 2024). In contrast, this study verifies that two distinct but complementary cognitive mechanisms, AI flow and perceived irreplaceability, jointly influence problematic GenAI use. The results show that the indirect effects via AI flow ($\beta = 0.113$) and perceived irreplaceability ($\beta = 0.120$) are both significant and do not differ significantly in magnitude (Diff = $-0.006$). These findings suggest that problematic GenAI use stems from a sense of loss of control arising from immersive use, as well as a sense of dependence triggered by indispensability. These two mechanisms, with proximate effects, provide a more refined operational theoretical model for future research.

## 5.3. Practical implications

The findings of this study have implications for design practice, education, organizational management, and social-level intervention strategies (Figure 3). First, at the level of AI system design, developers should consider the risk of losing control due to excessive immersion. The results show that AI flow is a key psychological mechanism driving problematic GenAI use. For a long time, product design has focused on user retention and interaction intensity. However, this study's results show that AI flow is a condition that leads to problematic GenAI use. Therefore, the core goal of AI products should not be to extend user usage time without limits, especially in contexts characterized by competition and social comparison. Interruption and reflection mechanisms should be added during high-intensity, immersive use. For example, when the system detects long periods of continuous interaction, instead of continuing to push information, non-intrusive feedback or stage-based task summaries can be provided to allow users to pause. Although such designs may reduce short-term activity, in the long run, they can enhance users' self-regulation and prevent immersive experiences from developing into psychological dependence.

Second, in education and organizational management, it is necessary to redefine the instrumental role of GenAI to address problematic use driven by perceived irreplaceability. The results show that the tool-based cognitive pathway promotes problematic use to a degree comparable to the immersion-based pathway. This shows that simply limiting how much people use GenAI is not enough, since the feeling that it cannot be replaced persists. Schools and institutions should focus on encouraging people to collaborate with AI rather than being replaced by it. People should not be judged solely on their AI capabilities. Training programs should help users see AI as a tool to aid their decision-making.

Finally, from a psychosocial intervention perspective, alleviating technology-related competitive anxiety is fundamental to the problem. Findings suggest that social comparison orientation and FoMO drive the use of GenAI. AI-driven performance evaluation should be taken with a grain of salt in the current social environment. Meanwhile, introducing multiple assessment criteria, such as originality, problem-definition ability, and integrative judgment, can effectively alleviate anxiety arising from comparisons of technologies.

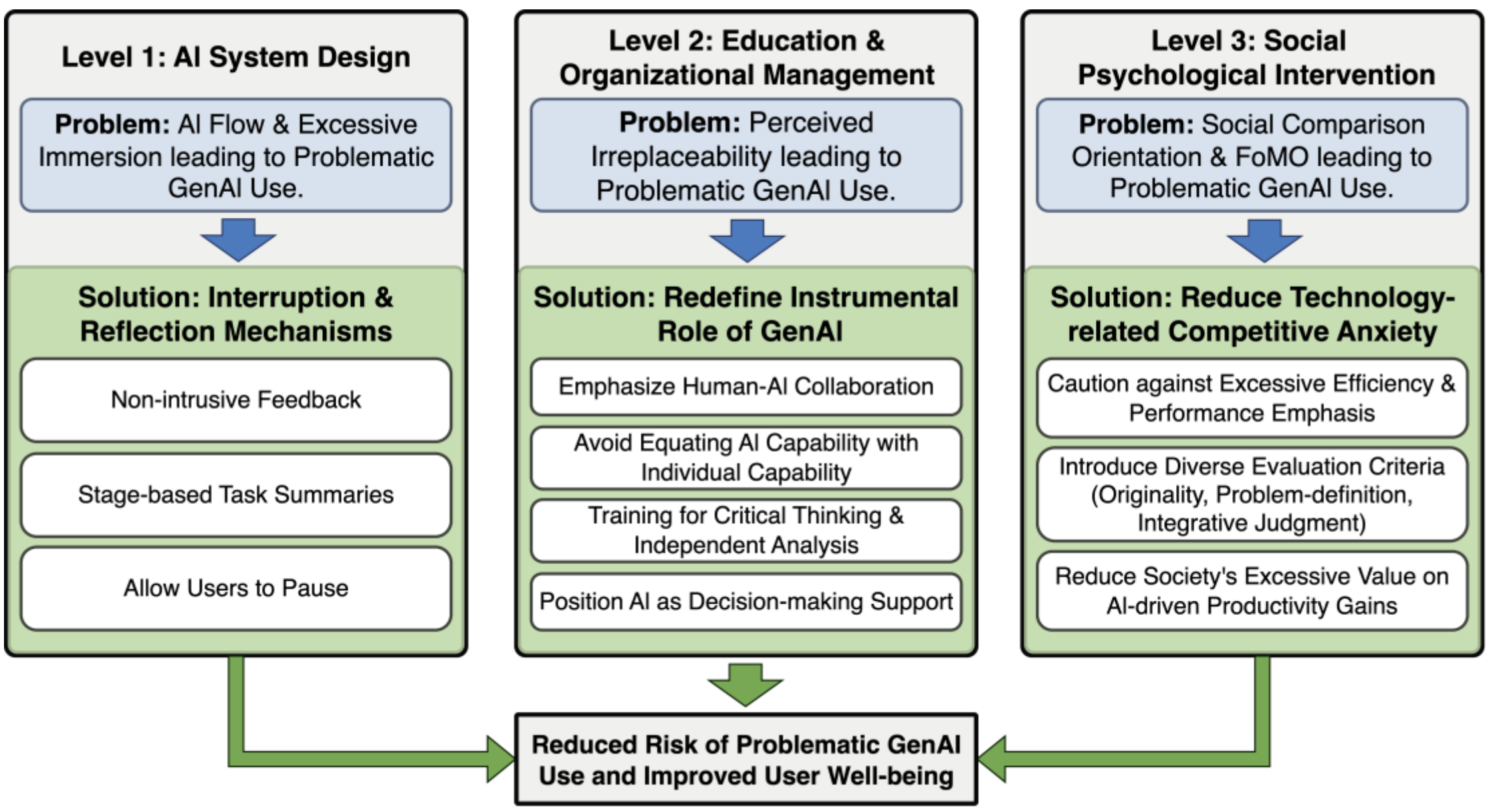

**Figure 3.** Management enlightenment.

## 6. Conclusion

This study examines the mechanisms by which social comparison tendencies influence problematic GenAI use. The research results show that social comparison orientation not only directly and significantly affects problematic GenAI use but also, through FoMO, affects AI flow and perceived Irreplaceability, resulting in significant indirect effects. These results indicate that platforms, organizations, and society should focus on problematic use to ensure continued development.

Although this study has produced specific theoretical outcomes, several limitations remain. First, although cross-sectional surveys are efficient for testing theoretical pathways, they are limited in their ability to explain causal effects (Du et al., 2025). Although this study examined the relationships among social comparison orientation, FoMO, AI flow, perceived irreplaceability, and problematic GenAI use, how these psychological states continuously reinforce one another over time remains unclear. It is also unknown at which point the pressure generated by social comparison orientation begins to transform into problematic use. These dynamic details are not yet known. Future research may adopt longitudinal tracking designs to restore these process mechanisms. Second, the data were collected using self-report questionnaires. Although robust statistical methods were applied, subjective and recall biases may still cause some interference. In particular, topics related to dependence are susceptible to social desirability bias. Therefore, future research may combine actual system usage records and psychological experimental indicators to conduct cross-validation and make the findings more objective. Finally, this study focuses on how people compare themselves to others, but users may also compare themselves to AI when using GenAI. The intense frustration people feel when they see their hard-earned skills easily surpassed by an AI's advanced capabilities may mark a new type of comparison between humans and AIs. It is unclear how this phenomenon relates to the FoMO mentality and how it affects user behavior. Incorporating human-AI comparisons into research models will help future researchers explore how technology can alter individuals' psychological mechanisms.

## Acknowledgement

This research received no specific grant from any funding agency in the public, commercial, or not-for-profit sectors.

## Declaration of Interest statement

The authors report there are no competing interests to declare.

## Data availability statement

The data used in this study are available from the corresponding author upon reasonable request.

## Ethical approval

This research was reviewed and approved by the Pusan National University Institutional Review Board (PNU IRB) and was classified as Exempt Review. The approval reference number is PNU IRB 2025-09-100. According to the exemption determination, written consent was waived because the study used a survey method that did not include personally identifiable information. Informed consent was obtained from all participants before data collection.

## Author contributions

Xuchao Zhang: Data curation, Formal analysis, Investigation, Methodology, Software, Validation, Visualization, Writing – original draft.
Jihye Lee: Conceptualization, Funding acquisition, Project administration, Resources, Supervision, Writing – review and editing.